\documentclass[amsmath,amssymb]{revtex4}
\usepackage{graphicx}
\usepackage{dcolumn}
\usepackage{bm}
\usepackage{color}

\newcommand*{\be}{\begin{equation}}
\newcommand*{\ee}{\end{equation}}

\begin{document}


\title{Squared Eigenfunctions for the Sasa-Satsuma Equation}
\author{Jianke Yang$^{1}$\footnote{Corresponding author, jyang@cems.uvm.edu},
        D. J. Kaup$^{2}$}
\affiliation{$^{1}$Department of Mathematics and Statistics,
University of Vermont, Burlington, VT 05401 \\
$^{2}$Department of Mathematics, University of Central Florida, Orlando, FL 32816 \\
}
\begin{abstract}
Squared eigenfunctions are quadratic combinations of Jost functions
and adjoint Jost functions which satisfy the linearized equation of
an integrable equation. They are needed for various studies related
to integrable equations, such as the development of its soliton
perturbation theory. In this article, squared eigenfunctions are
derived for the Sasa-Satsuma equation whose spectral operator is a
$3\times 3$ system, while its linearized operator is a $2\times 2$
system. It is shown that these squared eigenfunctions are sums of
two terms, where each term is a product of a Jost function and an
adjoint Jost function. The procedure of this derivation consists of
two steps: first is to calculate the variations of the potentials
via variations of the scattering data by the Riemann-Hilbert method.
The second one is to calculate the variations of the scattering data
via the variations of the potentials through elementary
calculations. While this procedure has been used before on other
integrable equations, it is shown here, for the first time, that for
a general integrable equation, the functions appearing in these
variation relations are precisely the squared eigenfunctions and
adjoint squared eigenfunctions satisfying respectively the
linearized equation and the adjoint linearized equation of the
integrable system. This proof clarifies this procedure and provides
a unified explanation for previous results of squared eigenfunctions
on individual integrable equations. This procedure uses primarily
the spectral operator of the Lax pair. Thus two equations in the
same integrable hierarchy will share the same squared eigenfunctions
(except for a time-dependent factor). In the Appendix, the squared
eigenfunctions are presented for the Manakov equations whose
spectral operator is closely related to that of the Sasa-Satsuma
equation.

\end{abstract}

\maketitle

\section{Introduction}
Squared eigenfunctions are quadratic combinations of Jost functions
and adjoint Jost functions which satisfy the linearized equation of
an integrable system. This name was derived from the fact that, for
many familiar integrable equations such as the KdV and NLS
equations, solutions of the linearized equations are squares of Jost
functions of the Lax pairs
\cite{Kaup1976a,Kaup1976b,Newell,massive_thirring,Benjamin,DNLS,
Herman}. Squared eigenfunctions are intimately related to the
integrable-equation theory. For instance, they are eigenfunctions of
the recursion operator of integrable equations \cite{AKNS,Fokas,
YangJMP,YangStud}. They also appear as self-consistent sources of
integrable equations \cite{Zeng,Hu}. A more important application of
squared eigenfunctions is in the direct soliton perturbation theory,
where squared eigenfunctions and their closure relation play a
fundamental role \cite{Kaup1990,Benjamin,DNLS,YangPRL03}.

Squared eigenfunctions have been derived for a number of integrable
equations including the KdV hierarchy, the AKNS hierarchy, the
derivative NLS hierarchy, the sine-Gordon equation, the
massive-Thirring model, the Benjamin-Ono equation, the matrix NLS
equations, the Kadomtsev-Petviashvili equation, etc. \cite{Fokas,
sine_Gordon, Benjamin, massive_thirring,YangJMP,YangStud,YangChen,
Gerdjikov}. Several techniques have been used in those derivations.
One is to notice the commutability relation between the linearized
operator and the recursion operator of the integrable equation
\cite{Fokas_commu_kdv, YangJMP,YangStud}. Thus squared
eigenfunctions are simply eigenfunctions of the recursion operator.
The drawback of this method is that one has to first derive the
recursion operator of the integrable equation and determine its
eigenfunctions, which can be highly non-trivial for many equations
(see \cite{sasa_recursion} for instance). Another technique is to
first calculate variations of the scattering data via the variations
of the potentials, then use the Wronskian relations to invert this
relation to get the variations of the potentials via variations of
the scattering data \cite{massive_thirring}. This Wronskian-relation
technique involves some ingenious steps and could run into
difficulties too if the problem is sufficiently complicated
\cite{massive_thirring}. The third technique is related to the
second one, except that one directly calculates the variations of
the potentials via variations of the scattering data by the
Riemann-Hilbert method \cite{massive_thirring,Benjamin}. This third
method is general and conceptually simpler. Regarding the second and
third methods however, one question which was never clarified is why
the functions appearing in the variation relations are indeed
squared eigenfunctions and adjoint squared eigenfunctions which
satisfy respectively the linearized equation and adjoint linearized
equation of an integrable system. In all known examples, this was
always found to be true by direct verifications. But whether and why
it would remain true for the general case was not known.

Recently, we were interested in the Sasa-Satsuma equation which is
relevant for the propagation of ultra-short optical pulses
\cite{sasa,Agrawal,Hasegawa}. This equation is integrable
\cite{sasa}. Two interesting features about this equation are that
its solitons are embedded inside the continuous spectrum of the
equation \cite{YangPRL99,YangPRL03}, and their shapes can be
double-humped for a wide range of soliton parameters \cite{sasa}.
One wonders how these double-humped solitons evolve when the
equation is perturbed --- a question which is interesting and
significant from both physical and mathematical points of view. This
question can be studied by a soliton perturbation theory. Due to the
embedded nature of these solitons, external perturbations will
generally excite continuous-wave radiation which is in resonance
with the soliton. So a key component in the soliton perturbation
theory would be to calculate this continuous-wave radiation, for
which squared eigenfunctions are needed (a similar situation occurs
in the soliton perturbation theory for the Hirota equation
\cite{YangPRL03}). A special feature of the Sasa-Satsuma equation is
that, while the linearized operator of this equation is $2\times 2$,
its spectral operator is $3\times 3$. How to build two-component
squared eigenfunctions from three-component Jost functions is an
interesting and curious question. As we shall see, only certain
components of the $3\times 3$ Jost functions and their adjoints are
used to construct the squared eigenfunctions and the adjoint squared
eigenfunctions of the Sasa-Satsuma equation. To calculate these
squared eigenfunctions for the Sasa-Satsuma equation, we have tried
to directly use the first method mentioned above. The recursion
operator for the Sasa-Satsuma equation has been derived recently
\cite{sasa_recursion}. However, that operator is quite complicated,
thus its eigenfunctions are difficult to obtain. The second method
mentioned above is met with difficulties too. Thus we choose the
third method for the Sasa-Satsuma equation and demonstrate here
certain advantages of this method.

In this paper, we further develop the third method, using the
Sasa-Satsuma equation as an example. We show that the functions
appearing in the expansion of the variations of the potentials are
always the squared eigenfunctions which satisfy the linearized
equation of an integrable system, and that the functions appearing
in the formulae for the variations of scattering data are always the
adjoint squared eigenfunctions which satisfy the adjoint linearized
equation of an integrable system. In addition, given these two
relations between the variations of the potentials and the
variations of scattering data, there naturally follows the closure
relation for the squared eigenfunctions and their adjoints, as well
as all the inner-product relations between the squared
eigenfunctions and their adjoints. Thus no longer is it necessary to
grind away at calculating these inner products from the asymptotics
of the Jost functions. Rather one can just read off the values of
the nonzero inner products from these two variation relations. This
clarifies the long-standing question regarding squared
eigenfunctions in connection with the linearized integrable
equation, and streamlines the third method as a general and
conceptually-simple procedure for the derivation of squared
eigenfunctions. We apply this method to the Sasa-Satsuma equation,
and find that it readily gives the squared eigenfunctions and
adjoint squared eigenfunctions.
The squared eigenfunctions for the Sasa-Satsuma equation are sums of
two terms, with each term being a product of a component of a Jost
function and a component of an adjoint Jost function. This two-term
structure of squared eigenfunctions is caused by the symmetry
properties of scattering data of the Sasa-Satsuma equation. It
should be noted that our derivation uses almost exclusively the
spectral operator of the Lax pair, thus all integrable equations
with the same spectral operator (such as the Sasa-Satsuma hierarchy)
will share the same set of squared eigenfunctions as we derived here
(except for a time-dependent factor which is equation-specific). An
additional benefit of the method we used is that for two integrable
equations with similar spectral operators, the derivation of their
squared eigenfunctions will be essentially the same. Thus one can
get squared eigenfunctions for one equation by minor modifications
for the other equation. As an example, we demonstrate in the
Appendix how squared eigenfunctions for the Manakov equations can be
easily obtained by minor modifications of our calculations for the
Sasa-Satsuma equation.

\section{The Riemann-Hilbert Problem}
 To start our analysis, we first formulate the Riemann-Hilbert problem for the
 Sasa-Satsuma equation which will be needed for later calculations.
 The Sasa-Satsuma equation is
 \be
 u_t+u_{xxx}+6|u|^2u_x+3u(|u|^2)_x=0.
 \ee
 Its spectral (scattering) problem of the Lax pair is \cite{sasa}
 \begin{equation} \label{spectr}
 Y_x=-i\zeta\Lambda Y+QY,
 \end{equation}
 where $Y$ is a matrix function, $\Lambda=\mathrm{diag}(1,1,-1)$,
 \be  \label{Q}
 Q=\left(\begin{array}{ccc}0 & 0 & u \\
 0 & 0 & u^* \\
 -u^* & -u & 0 \end{array}\right)
 \ee
 is the potential matrix, the superscript ``*" represents complex conjugation,
 and $\zeta$ is a spectral parameter.
 In this paper, we always assume that the potential $u(x)$ decays to zero sufficiently
 fast as $x\to \pm \infty$.
 Notice that this matrix $Q$ has two symmetry properties. One is that it is
 anti-Hermitian, i.e. $Q^\dagger=-Q$,
 where the superscript ``$\dagger$" represents the Hermitian of a
 matrix. The other one is that
 \be   \label{Qsym}
 \sigma Q \sigma=Q^*,  \qquad \mathrm{where} \quad
 \sigma=\sigma^{-1}=\left(\begin{array}{ccc}0 & 1 & 0 \\
 1 & 0 & 0 \\
 0 & 0 & 1 \end{array}\right).
 \ee
 Introducing new variables
 \be
 J=YE^{-1}, \qquad E=e^{-i\zeta\Lambda x},
 \ee
 then Eq. (\ref{spectr}) becomes
 \begin{equation} \label{Jx}
 J_x=-i\zeta[\Lambda, J]+QJ,
 \end{equation}
 where $[\Lambda, J]=\Lambda J-J\Lambda$.
 The matrix Jost solutions $J_\pm(x,\zeta)$ of Eq. (\ref{Jx}) are defined by the
 asymptotics
 \be \label{Jasym}
 J_\pm\to\openone, \quad \mathrm{as} \quad x \to \pm\infty,
 \ee
 where ``$\openone$" is the unit matrix. Here the subscripts in $J_\pm$ refer to which end of the $x$-axis the
 boundary conditions are set.
 Since $\mathrm{tr}Q=0$, using Abel's formula on Eq. (\ref{Jx}), we
 see that $\det J_\pm=1$ for all $x$. In addition, since $J_{\pm}E$ are both
 solutions of the linear equations (\ref{spectr}), they are not independent and are linearly related
 by the scattering matrix $S(\zeta)=[s_{ij}(\zeta)]$:
 \be\label{scatt}
 J_-=J_+ESE^{-1},  \qquad \det S=1
 \ee
 for real $\zeta$, i.e. $\zeta \in \mathbb{R}$ (for non-real $\zeta$, certain elements in
 $S$ may not be well defined).

 Due to the two symmetry properties of the potential matrix $Q$,
 the Jost solutions $J_{\pm}$ and the scattering matrix $S$ satisfy
 the corresponding symmetry properties. One is the
 involutive property. Since the potential $Q$ is
 anti-Hermitian, we see that $J_\pm^\dag(\zeta^*)$ and
 $J_\pm^{-1}(\zeta)$ both satisfy the adjoint equation of
 (\ref{Jx}). In addition, we see from Eq. (\ref{Jasym}) that $J_\pm^\dag(\zeta^*)$ and
 $J_\pm^{-1}(\zeta)$ have the same large-$x$ asymptotics, thus they
 are equal to each other:
 \be\label{involJ}
 J_\pm^\dag(\zeta^*)=J_\pm^{-1}(\zeta).
 \ee
 Then in view of Eq. (\ref{scatt}), we see that
 \be\label{involS}
 S^\dag(\zeta^*)=S^{-1}(\zeta).
 \ee
 To derive the other symmetry properties of the
 Jost solutions and the scattering matrix, we notice that due to the
 symmetry (\ref{Qsym}) of the potential $Q$, it is easy to see that $\sigma J_{\pm}^*(-\zeta^*)\sigma$
 also satisfies Eq. (\ref{Jx}). Then in view of the asymptotics (\ref{Jasym}), we find that
 Jost solutions possesses the following additional symmetry
 \be \label{Jaddition}
 J_{\pm}(\zeta)=\sigma J_{\pm}^*(-\zeta^*)\sigma.
 \ee
 From this symmetry and the relation (\ref{scatt}), we see that the scattering
 matrix $S$ possesses the additional symmetry
 \be  \label{Saddition}
 S(\zeta)=\sigma S^*(-\zeta^*)\sigma.
 \ee

 Analytical properties of the Jost solutions play a fundamental role in the
 Riemann-Hilbert formulation for the scattering problem (\ref{Jx}).
 For convenience, we express $J_{\pm}$ as:
 \be  \label{Phidef}
 J_{-}=\Phi E^{-1}, \quad \Phi=[\phi_1, \phi_2, \phi_3],
 \ee
 \be
 J_{+}=\Psi E^{-1}, \quad \Psi=[\psi_1, \psi_2, \psi_3],
 \ee
 where $\Phi$ and $\Psi$ are fundamental solutions of the original spectral problem
 (\ref{spectr}) and are related by the scattering matrix as
 \be \label{PhiPsiS}
 \Phi=\Psi S.
 \ee
 Then in view of the boundary conditions (\ref{Jasym}),
 the spectral equation (\ref{Jx}) for $J_{\pm}$ can be rewritten into
 the following Volterra-type integral equations:
 \be \label{J-int}
 J_-(\zeta; x)=\openone +\int_{-\infty}^x e^{i\zeta \Lambda
 (y-x)}Q(y)J_-(\zeta; y)e^{i\zeta \Lambda (x-y)}dy,
 \ee
 \be \label{J+int}
 J_+(\zeta; x)=\openone -\int^{\infty}_x e^{i\zeta \Lambda
 (y-x)}Q(y)J_-(\zeta; y)e^{i\zeta \Lambda (x-y)}dy.
 \ee
 These integral equations always have solutions when the integrals
 on their right hand sides converge. Due to the structure (\ref{Q}) of the potential $Q$, we
 easily see that Eq. (\ref{J-int}) for the first and second columns of $J_{-}$
 contain only the exponential factor $e^{i\zeta (x-y)}$ which decays
 when $\zeta$ is in the upper half plane $\mathbb{C}_+$, and Eq. (\ref{J+int})
 for the third column of
 $J_+$ contains only the exponential factor $e^{i\zeta (y-x)}$ which
 also falls off for $\zeta \in \mathbb{C}_+$. Thus these three
 columns can be analytically extended to $\zeta \in \mathbb{C}_+$. In other words,
 Jost solutions
 \be \label{P+}
 P_+=[\phi_1, \phi_2, \psi_3]e^{i\zeta \Lambda
 x}=J_{-}H_1+J_+H_2
 \ee
 are analytic in $\zeta \in \mathbb{C}_+$, where
 \be
 H_1=\mathrm{diag}(1,1,0), \qquad H_2=\mathrm{diag}(0,0,1).
 \ee
 Here the subscript in $P$ refers to which half plane the functions are analytic
 in. From the Volterra integral equations for $P_+$, we see that
 \be  \label{P+asym}
 P_+(x,\zeta)\to\openone \quad \mathrm{as} \quad \zeta \in \mathbb{C}_+ \to
 \infty.
 \ee
 Similarly, Jost functions
 $[\psi_1, \psi_2, \phi_3]e^{i\zeta \Lambda x}$ are analytic in
 $\zeta \in \mathbb{C}_-$, and their large-$\zeta$ asymptotics is
 \be  \label{asym2a}
 [\psi_1, \psi_2, \phi_3]e^{i\zeta \Lambda x} \to\openone \quad \mathrm{as} \quad \zeta \in \mathbb{C}_- \to
 \infty.
 \ee

 To obtain the analytic counterpart of $P_+$ in $\mathbb{C}_-$, we
 consider the adjoint spectral equation of (\ref{Jx}):
 \be\label{adj}
 K_x=-i\zeta[\Lambda,K]-K Q.
 \ee
 The inverse matrices $J_{\pm}^{-1}$ satisfy this adjoint equation.
 Notice that
 \be
 J_{-}^{-1}=E\Phi^{-1}, \quad J_{+}^{-1}=E\Psi^{-1}.
 \ee
 Let us express $\Phi^{-1}$ and $\Psi^{-1}$ as a collection of rows,
 \be  \label{Phiinv}
 \Phi^{-1}\equiv \bar{\Phi} =\left[\begin{array}{c}\bar{\phi}_1 \\
 \bar{\phi}_2 \\ \bar{\phi}_3 \end{array}\right], \qquad
 \Psi^{-1}\equiv \bar{\Psi} =\left[\begin{array}{c}\bar{\psi}_1 \\
 \bar{\psi}_2 \\ \bar{\psi}_3 \end{array}\right],
 \ee
 where the overbar refers to the adjoint quantity,
 then by similar techniques as used above, we can show that the first
 and second rows of $J_{-}^{-1}$ and the third row of $J_{+}^{-1}$
 are analytic in $\zeta \in \mathbb{C}_-$, i.e. adjoint Jost solutions
 \be  \label{P-}
 P_-=e^{-i\zeta \Lambda x} \left[\begin{array}{c}\bar{\phi}_1 \\
 \bar{\phi}_2 \\ \bar{\psi}_3 \end{array}\right]=
 H_1J_{-}^{-1}+H_2J_+^{-1}
 \ee
 are analytic in $\zeta \in \mathbb{C}_-$. In addition, their large-$\zeta$
 asymptotics is
 \be  \label{P-asym}
 P_-(x,\zeta)\to\openone \quad \mathrm{as} \quad \zeta \in \mathbb{C}_- \to
 \infty.
 \ee
 Similarly, the first
 and second rows of $J_{+}^{-1}$ and the third row of $J_{-}^{-1}$,
 i.e. $e^{-i\zeta x}\bar{\psi}_1$, $e^{-i\zeta x}\bar{\psi}_2$, and
 $e^{i\zeta x}\bar{\phi}_3$,
 are analytic in $\zeta \in \mathbb{C}_+$, and their large-$\zeta$
 asymptotics is
 \be  \label{asym2b}
 e^{-i\zeta \Lambda x}\left[\begin{array}{c}\bar{\psi}_1 \\
 \bar{\psi}_2 \\ \bar{\phi}_3 \end{array}\right] \to \openone
 \quad \mathrm{as} \quad \zeta \in \mathbb{C}_+ \to
 \infty.
 \ee
 In view of the involution properties (\ref{involJ}) of $J_{\pm}$,
 we see that the analytic solutions $P_{\pm}$ satisfy
 the involutive property as well:
 \be\label{invol}
 P_+^\dag(\zeta^*)=P_-(\zeta).
 \ee
 This property can be taken as a definition of the analytic function
 $P_-$ from the known analytic function $P_+$.

 The analytic properties of Jost functions described above have
 immediate implications on the analytic properties of the scattering
 matrix $S$.
 Let us denote
 \be \label{Sinv}
 S^{-1}(\zeta)\equiv \bar{S}(\zeta)=[\bar{s}_{ij}(\zeta)],
 \ee
 then since
 \be
 S=\Psi^{-1}\Phi=\left[\begin{array}{c}\bar{\psi}_1 \\
 \bar{\psi}_2 \\ \bar{\psi}_3 \end{array}\right] [\phi_1, \phi_2,
 \phi_3], \qquad
 \bar{S}=\Phi^{-1}\Psi=\left[\begin{array}{c}\bar{\phi}_1 \\
 \bar{\phi}_2 \\ \bar{\phi}_3 \end{array}\right] [\psi_1, \psi_2,
 \psi_3],
 \ee
 we see immediately that $s_{11}, s_{12}, s_{21}, s_{22}$ and
 $\bar{s}_{33}$ can be analytically extended to the upper half plane $\zeta\in \mathbb{C}_+$,
 while $\bar{s}_{11}, \bar{s}_{12}, \bar{s}_{21}, \bar{s}_{22}$ and
 $s_{33}$ can be analytically extended to the lower half plane $\zeta\in
 \mathbb{C}_-$. In addition, their large-$\zeta$ asymptotics are
 \be  \label{sijasym1}
 \left(\begin{array}{cc} s_{11} & s_{12} \\ s_{21}
 & s_{22}\end{array}\right) \to \openone, \quad \bar{s}_{33}
 \to 1, \quad \mathrm{as} \quad \zeta \in \mathbb{C}_+ \to
 \infty,
 \ee
 and
 \be  \label{sijasym2}
 \left(\begin{array}{cc} \bar{s}_{11} & \bar{s}_{12} \\ \bar{s}_{21}
 & \bar{s}_{22}\end{array}\right) \to \openone, \quad s_{33}
 \to 1, \quad \mathrm{as} \quad \zeta \in \mathbb{C}_- \to
 \infty.
 \ee

 Hence we have constructed two matrix functions $P_+$ and
 $P_-$ which are analytic in $\mathbb{C}_+$ and $\mathbb{C}_-$
 respectively. On the real line, using Eqs. (\ref{scatt}),
 (\ref{P+}), (\ref{P-}), and (\ref{Sinv}), we easily see that
 \be\label{RHP}
 P_-(\zeta)P_+(\zeta)=G(\zeta), \qquad \zeta \in \mathbb{R},
 \ee
 where
 \be \label{G}
 G=E(H_1+H_2S)(H_1+S^{-1}H_2)E^{-1}=E\left(\begin{array}{ccc} 1 & 0 &
 \bar{s}_{13} \\ 0 & 1 & \bar{s}_{23} \\ s_{31} & s_{32} & 1
 \end{array}\right)E^{-1}.
 \ee
 Equation (\ref{RHP}) determines a matrix Riemann-Hilbert problem.
 The normalization condition for this Riemann-Hilbert problem can be
 seen from (\ref{P+asym}) and (\ref{P-asym}) as
 \be\label{normal}
 P_\pm(x,\zeta)\to\openone \quad \mathrm{as} \quad \zeta\in \mathbb{C}_\pm \to
 \infty.
 \ee
 If this problem can be solved, then
 the potential $Q$ can be reconstructed from an asymptotic expansion
 of its solution for large $\zeta$. Indeed, writing $P_+$ as
 \be \label{Pexpand}
 P_+(x,\zeta)=\openone+\zeta^{-1}P^{(1)}(x)+\zeta^{-2}P^{(2)}(x)+{\cal O}(\zeta^{-3}),
 \ee
 inserting it into Eq. (\ref{Jx}) and comparing terms of the same order in $\zeta^{-1}$,
 we find at $O(1)$ that
 \be \label{reconst}
 Q=i[\Lambda,P^{(1)}].
 \ee
 Thus the potential $Q$ can be reconstructed from $P^{(1)}$. At
 $O(\zeta^{-1})$, we find that
 \be
 P^{(1)}_x=-i[\Lambda, P^{(2)}]+QP^{(1)}.
 \ee
 From the above two equations as well as the large-$x$ asymptotics
 of $P_+(x,\zeta)$ from Eq. (\ref{Jasym}), we see that the full
 matrix $P^{(1)}$ is
 \be
 P^{(1)}(x)=\frac{1}{2i}\left[\begin{array}{ccc}\int_{-\infty}^x |u(y)|^2 dy & \quad \int_{-\infty}^x u^2(y) dy & u(x)
 \\   \\
 \int_{-\infty}^x u^{*2}(y) dy & \quad \int_{-\infty}^x |u(y)|^2 dy & u^*(x)
 \\  \\
 u^*(x) & u(x) & 2\int_x^{\infty} |u(y)|^2 dy \end{array}\right].
 \ee
 This matrix, together with Eq. (\ref{Pexpand}), gives the
 leading-order asymptotic expansions for the analytical functions $P_+$.
 Leading-order asymptotic expansions for $P_-$ can be similarly
 derived.

 Elements of the scattering matrix $S$ are not all independent.
 Indeed, since the potential matrix $Q$ contains only one
 independent function $u(x)$, matrix $S$ should contain only one
 independent element as well. To get
 the dependence of scattering coefficients, we first notice from equation
 $S\bar{S}=\openone$ that
 \be  \label{sRH}
 s_{33}(\zeta)\bar{s}_{33}(\zeta)=1-s_{31}(\zeta)\bar{s}_{13}(\zeta)-s_{32}(\zeta)
 \bar{s}_{23}(\zeta), \quad \zeta \in \mathbb{R}.
 \ee
 Since $s_{33}$ and $\bar{s}_{33}$ are analytic in $\mathbb{C}_-$
 and $\mathbb{C}_+$ respectively, the above equation defines a
 (scalar) Riemann-Hilbert problem for $(s_{33}, \bar{s}_{33})$,
 under the canonical normalization conditions which can be seen from
 (\ref{sijasym1})-(\ref{sijasym2}). Thus
 $(s_{33}, \bar{s}_{33})$ can be uniquely determined by the
 locations of their zeros in $\mathbb{C}_\pm$ together with the scattering data
 $(s_{31}, s_{32}, \bar{s}_{13}, \bar{s}_{23})$ on the real line $\zeta \in \mathbb{R}$
 by the Plemelj formula. Similarly, from equation $S\bar{S}=\openone$ we have
 \be
 \left(\begin{array}{cc} \bar{s}_{11} & \bar{s}_{12} \\ \bar{s}_{21}
 & \bar{s}_{22}\end{array}\right)
 \left(\begin{array}{cc} s_{11} & s_{12} \\ s_{21}
 & s_{22}\end{array}\right)=\openone - \left(\begin{array}{c} \bar{s}_{13} \\ \bar{s}_{23}
 \end{array}\right)(s_{31}, s_{32}), \quad \zeta \in \mathbb{R},
 \ee
 which defines another (matrix) Riemann-Hilbert problem for $(\bar{s}_{11}, \bar{s}_{12}, \bar{s}_{21},
 \bar{s}_{22})$ and $(s_{11}, s_{12}, s_{21}, s_{22})$, under the
 canonical normalization conditions which can be seen from
 (\ref{sijasym1})-(\ref{sijasym2}). Thus these
 analytical scattering elements can be determined by the scattering data
 $(s_{31}, s_{32}, \bar{s}_{13}, \bar{s}_{23})$ on the real line $\zeta \in
 \mathbb{R}$ as well. The scattering data $(s_{31}, s_{32}, \bar{s}_{13},
 \bar{s}_{23})$ themselves are dependent on each other by the
 symmetry relations (\ref{involS}) and (\ref{Saddition}).
 Specifically, for $\xi\in \mathbb{R}$,
 $\bar{s}_{13}(\xi)=s_{31}^*(\xi), \bar{s}_{23}(\xi)=s_{32}^*(\xi)$,
 and $s_{32}(\xi)=s_{31}^*(-\xi)$. Thus the scattering matrix
 $S$ indeed contains only a single independent element as we would
 expect.

 In the next section, we will derive squared eigenfunctions for the Sasa-Satsuma equation.
 For the convenience of that derivation, we introduce the following notations:
 \be \label{rou}
 \rho_1\equiv \frac{s_{31}}{s_{33}}, \quad
 \rho_2\equiv \frac{s_{32}}{s_{33}}, \quad
 \bar{\rho}_1\equiv \frac{\bar{s}_{13}}{\bar{s}_{33}}, \quad
 \bar{\rho}_2\equiv \frac{\bar{s}_{23}}{\bar{s}_{33}}.
 \ee
 Due to the
 symmetry conditions (\ref{involS}) and (\ref{Saddition}) of the scattering matrix $S$,
 we see that $\rho_k$ and $\bar{\rho}_k$ satisfy the symmetry conditions
 \be \label{rho_sym}
 \bar{\rho}_k(\zeta)=\rho^*_k(\zeta), k=1, 2, \quad
 \bar{\rho}_1(\zeta)=\rho_2(-\zeta), \quad
 \bar{\rho}_2(\zeta)=\rho_1(-\zeta), \quad \zeta \in \mathbb{R}.
 \ee
 In the next section, symmetry properties of $\Phi$ and $\Psi$ will
 also be needed.
 Due to the symmetry conditions (\ref{involJ}) and (\ref{Jaddition})
 of the Jost solutions, $\Phi$ and $\Psi$ satisfy the
 symmetry conditions
 \be \label{involPhi}
 \Phi^\dagger(\zeta^*)=\Phi^{-1}(\zeta), \quad \Psi^\dagger(\zeta^*)=\Psi^{-1}(\zeta),
 \ee
 and
 \be \label{Phiaddition}
 \Phi_{\pm}(\zeta)=\sigma \Phi_{\pm}^*(-\zeta^*)\sigma, \quad
 \Psi_{\pm}(\zeta)=\sigma \Psi_{\pm}^*(-\zeta^*)\sigma.
 \ee
 Symmetry (\ref{involPhi}) means that
 \be  \label{involphi}
 \phi_k^\dagger(\zeta^*)=\bar{\phi}_k(\zeta),
 \quad \psi_k^\dagger(\zeta^*)=\bar{\psi}_k(\zeta), \quad k=1, 2, 3,
 \ee
 while symmetry (\ref{Phiaddition}), together with (\ref{involphi}), means that
 \be  \label{phi_sym}
 \sigma \phi_1(\zeta)=\bar{\phi}_2^T(-\zeta), \quad
 \sigma \phi_2(\zeta)=\bar{\phi}_1^T(-\zeta), \quad
 \sigma \phi_3(\zeta)=\bar{\phi}_3^T(-\zeta).
 \ee
 Here the superscript "$T$" represents the transpose of a matrix.
 These symmetry properties will be important for deriving
 the final expressions of squared eigenfunctions for the
 Sasa-Satsuma equation in the next section.

\section{Squared eigenfunctions and their closure relation}
 In this section, we calculate the variation of the potential via
 variations of the scattering data, then calculate
 variations of the scattering data via the variation of the
 potential. The first step will yield squared eigenfunctions, and it
 will be done by the Riemann-Hilbert
 method. The second step will yield adjoint squared eigenfunctions, and
 it will be done using basic relations of the
 spectral problem (\ref{spectr}). For the ease of
 presentation, we first assume that $s_{33}$ and $\bar{s}_{33}$ have
 no zeros in their respective planes of analyticity, i.e. the
 spectral problem (\ref{spectr}) has no discrete eigenvalues. This
 facilitates the derivation of squared eigenfunctions. The
 results for the general case of $s_{33}$ and $\bar{s}_{33}$ having
 zeros will be given in the end of this section.

\subsection{Variation of the potential and squared eigenfunctions}

 In this subsection, we derive the variation of the potential via
 variations of the scattering data, which will readily yield
 the squared eigenfunctions satisfying the linearized Sasa-Satsuma equation.
 Our derivation will be based on the Riemann-Hilbert
 method.

 To proceed, we define the following matrix functions
 \be \label{Fdef}
 F_+=P_+ \: \mbox{diag}(1, 1, \frac{1}{\bar{s}_{33}}), \qquad
 F_-=P_-^{-1}\: \mbox{diag}(1, 1, s_{33}).
 \ee
 The reason to introduce diagonal matrices with $s_{33}$ and
 $\bar{s}_{33}$ in $F_{\pm}$ is to obtain a new Riemann-Hilbert problem (\ref{RHF}) with
 a connection matrix $\tilde{G}$ which depends on $\rho_k$ and
 $\bar{\rho}_k$ rather than $s_{ij}$ and $\bar{s}_{ij}$. This way, the variation of the
 potential will be expressed in terms of variations in $\rho_k$ and
 $\bar{\rho}_k$.  When $\bar{s}_{33}$ and $s_{33}$ have
 no zeros in $\mathbb{C}_+$ and $\mathbb{C}_-$ respectively,
 then $F_{\pm}$ as well as $F_{\pm}^{-1}$ are analytic in $\mathbb{C}_{\pm}$.
 On the real line, they are related by
 \be \label{RHF}
 F_+(\zeta)=F_-(\zeta) \tilde{G}(\zeta), \qquad \zeta \in \mathbb{R},
 \ee
 where
 \be \label{Gtilde}
 \tilde{G}=\mbox{diag}(1, 1, \frac{1}{s_{33}})\: G \: \mbox{diag}(1, 1, \frac{1}{\bar{s}_{33}})
 =E\left(\begin{array}{lll} 1 & 0 &
 \bar{\rho}_{1} \\ 0 & 1 & \bar{\rho}_{2} \\ \rho_{1} & \rho_{2} &
 1+\rho_1\bar{\rho}_1+\rho_2\bar{\rho}_2
 \end{array}\right)E^{-1}.
 \ee
 Here the relation (\ref{sRH}) has been used. Eq. (\ref{RHF}) defines
 a regular Riemann-Hilbert problem (i.e. without zeros).

 Next we take the variation of the Riemann-Hilbert problem
 (\ref{RHF}), and get
 \be
 \delta F_+=\delta F_-\tilde{G}+F_-\delta \tilde{G}, \qquad \zeta \in \mathbb{R}.
 \ee
 Utilizing Eq. (\ref{RHF}), we can rewrite the above equation as
 \be  \label{RHdF}
 \delta F_+ F_+^{-1} = \delta F_- F_-^{-1}+F_-\delta
 \tilde{G}F_+^{-1}, \qquad \zeta \in \mathbb{R},
 \ee
 which defines yet another regular Riemann-Hilbert problem for $\delta F
 F^{-1}$. Unlike the previous Riemann-Hilbert problems (\ref{RHP})
 and (\ref{RHF}) which were in matrix product forms, the present Riemann-Hilbert
 problem (\ref{RHdF}) can be explicitly solved. Using the Plemelj
 formula, the general solution of this Riemann-Hilbert problem is
 \be \label{deltaF}
 \delta F F^{-1}(\zeta; x)=A_0(x)+\frac{1}{2\pi i}
 \int_{-\infty}^\infty \frac{\Pi(\xi; x)}{\xi-\zeta} d\xi,
 \ee
 where
 \be \label{Pidef}
 \Pi(\xi; x) \equiv F_-(\xi; x)\: \delta \tilde{G}(\xi; x)\: F_+^{-1}(\xi;
 x), \qquad \xi \in \mathbb{R}.
 \ee
 Now we consider the large-$\zeta$ asymptotics of this solution. Notice from expansions (\ref{Pexpand}),
 (\ref{sijasym1}) and relation (\ref{Fdef}) that as $\zeta \to
 \infty$,
 \be  \label{Fasym}
 F_+(\zeta; x) \to
 \left[\begin{array}{ccc} 1+O(\frac{1}{\zeta}) & O(\frac{1}{\zeta})
 & \frac{u(x)}{2i\zeta} \\  \\
 O(\frac{1}{\zeta}) & 1+O(\frac{1}{\zeta})& \frac{u^*(x)}{2i\zeta}
 \\  \\
 \frac{u^*(x)}{2i\zeta} & \frac{u(x)}{2i\zeta} & 1+O(\frac{1}{\zeta})
 \end{array}\right],
 \qquad
 \delta F_+(\zeta; x) \to
 \left[\begin{array}{ccc} O(\frac{1}{\zeta}) & O(\frac{1}{\zeta})
 & \frac{\delta u(x)}{2i\zeta} \\  \\
 O(\frac{1}{\zeta}) & O(\frac{1}{\zeta})& \frac{\delta u^*(x)}{2i\zeta}
 \\  \\
 \frac{\delta u^*(x)}{2i\zeta} & \frac{\delta u(x)}{2i\zeta} & O(\frac{1}{\zeta}) \end{array}\right].
 \ee
 In addition, it is easy to see that
 \be
 \int_{-\infty}^\infty \frac{\Pi(\xi; x)}{\xi-\zeta} d\xi
 \longrightarrow
 -\frac{1}{\zeta} \int_{-\infty}^\infty \Pi(\xi; x)d\xi, \qquad \zeta
 \to \infty.
 \ee
 When these large-$\zeta$ expansions are substituted into Eq.
 (\ref{deltaF}), at $O(1)$, we find that $A_0(x)=0$. At
 $O(\zeta^{-1})$, we get
 \be \label{du}
 \delta u(x) =-\frac{1}{\pi} \int_{-\infty}^\infty \Pi_{13}(\xi;
 x)d\xi, \qquad
 \delta u^*(x) =-\frac{1}{\pi} \int_{-\infty}^\infty \Pi_{31}(\xi;
 x)d\xi.
 \ee
 Now we calculate the elements $\Pi_{13}$ and $\Pi_{31}$. From Eq. (\ref{Gtilde}), we see that
 \be \label{dGtilde}
 \delta\tilde{G}=E\left(\begin{array}{ccc} 0 & 0 &
 \delta\bar{\rho}_{1} \\ 0 & 0 & \delta\bar{\rho}_{2} \\ \delta\rho_{1} & \delta\rho_{2} &
 \rho_1\delta\bar{\rho}_1+\bar{\rho}_1\delta\rho_1+
 \rho_2\delta\bar{\rho}_2+\bar{\rho}_2\delta\rho_2
 \end{array}\right)E^{-1}.
 \ee
 Then in view of the $F_{\pm}$ definitions (\ref{Fdef}) and
 $P_{\pm}$ expressions (\ref{P+}), (\ref{P-}), we find from (\ref{Pidef}) that
 \be
 \Pi(\zeta; x)=J_-E (H_1+H_2S)^{-1}\mbox{diag}(1, 1,
 s_{33})E^{-1}\: \delta\tilde{G}\: E\: \mbox{diag}(1, 1,
 \bar{s}_{33})(H_1+S^{-1}H_2)^{-1}E^{-1}J_-^{-1},
 \ee
 which simplifies to
 \be \label{Pi}
 \Pi(\zeta; x)=\Phi \left(\begin{array}{ccc} 0 & 0 & \delta\bar{\rho}_{1}
 \\ 0 & 0 & \delta\bar{\rho}_{2} \\
 \delta\rho_{1} & \delta\rho_{2} & 0
 \end{array}\right) \bar{\Phi}.
 \ee
 This relation, together with Eq. (\ref{du}), shows that the variation
 of the potential $\delta u$ can be expanded into quadratic combinations
 between Jost solutions $\Phi$ and adjoint Jost solutions $\bar{\Phi}$.
 This relation is generic in integrable systems.

 Now we calculate explicit expressions for $\delta u$. Inserting
 (\ref{Pi}) into (\ref{du}) and recalling our notations
 (\ref{Phiinv}), we readily find that
 \be  \label{du4term}
 \delta u =-\frac{1}{\pi} \int_{-\infty}^\infty \left(
 \phi_{31}\bar{\phi}_{13}\delta \rho_1 +
 \phi_{31}\bar{\phi}_{23}\delta \rho_2 +
 \phi_{11}\bar{\phi}_{33}\delta \bar{\rho}_1+
 \phi_{21}\bar{\phi}_{33}\delta \bar{\rho}_2\right)
 d\xi.
 \ee
 Here the notations are
 \be  \label{phik}
 \phi_k=\left[\begin{array}{c} \phi_{k1} \\ \phi_{k2} \\ \phi_{k3}
 \end{array}\right], \quad
 \psi_k=\left[\begin{array}{c} \psi_{k1} \\ \psi_{k2} \\ \psi_{k3}
 \end{array}\right], \quad
 \bar{\phi}_k=\left[\bar{\phi}_{k1},  \bar{\phi}_{k2},  \bar{\phi}_{k3}
 \right], \quad
 \bar{\psi}_k=\left[\bar{\psi}_{k1},  \bar{\psi}_{k2},  \bar{\psi}_{k3}
 \right], \quad k=1, 2, 3.
 \ee
 Utilizing the symmetry relations (\ref{rho_sym}) and (\ref{phi_sym}),
 the above $\delta u$ formula reduces to
 \be
 \delta u =-\frac{1}{\pi} \int_{-\infty}^\infty \left[\left(
 \phi_{31}\bar{\phi}_{13}+\phi_{33}\bar{\phi}_{12}\right)\delta \rho_1 +
 \left(\phi_{11}\bar{\phi}_{33}+\phi_{13}\bar{\phi}_{32}\right)\delta \bar{\rho}_1\right]
 d\xi.
 \ee
 This is an important step in our derivation, where symmetry
 conditions play an important role. Regarding $\delta u^*$, its
 formula can be obtained by taking the
 complex conjugate of the above equation and simplified
 by using the symmetry relations (\ref{involphi}). Defining
 functions
 \be  \label{Z12def}
 Z_1=\left[\begin{array}{c}
 \phi_{31}\bar{\phi}_{13}+\phi_{33}\bar{\phi}_{12} \\
 \phi_{32}\bar{\phi}_{13}+\phi_{33}\bar{\phi}_{11}
 \end{array}\right], \quad
 Z_2=\left[\begin{array}{c}
 \phi_{11}\bar{\phi}_{33}+\phi_{13}\bar{\phi}_{32} \\
 \phi_{12}\bar{\phi}_{33}+\phi_{13}\bar{\phi}_{31}
 \end{array}\right],
 \ee
 then the final expression for the variation of the potential $(\delta u,
 \delta u^*)^T$ is
 \be  \label{deltau}
 \left[\begin{array}{c}
 \delta u(x) \\
 \delta u^*(x) \end{array}\right]=
 -\frac{1}{\pi} \int_{-\infty}^\infty  \left[Z_1(\xi; x) \delta \rho_1(\xi)+Z_2(\xi; x)
 \delta \bar{\rho}_1(\xi) \right]d\xi.
 \ee
 Notice here that due to the symmetry relations
 (\ref{involphi}), $Z_2(\zeta)$ is equal to $Z_1^*(\zeta^*)$ with
 its two components swapped. Also notice that $Z_1$ and $Z_2$ are
 the sum of two terms, where each term is a product of a component
 of a Jost function and a component of an adjoint Jost function.
 The two-term feature of these functions is caused by the symmetry properties of the scattering data and Jost
 functions, which enable us to combine terms together in the general expansion
 (\ref{du4term}). In the massive Thirring model, the counterparts of
 functions $Z_{1, 2}$ are also sums of two terms \cite{massive_thirring}.
 The two-term structure there is not due to symmetry properties, but
 rather due to the asymptotics of its Jost functions in the spectral plane.
 The feature of each term in $Z_1$ and $Z_2$ being a product between a Jost
 function and an adjoint Jost function, on the other hand, is a
 generic feature in integrable systems. In previous studies on many
 integrable equations (such as the KdV, NLS, derivative NLS, and
 massive Thirring equations), it was found that these functions were often
 ``squares" or products of Jost functions themselves (thus the name ``squared eigenfunctions")
 \cite{Kaup1976a,Newell,YangChen,massive_thirring}. This was so simply
 because the spectral operators in those systems were
 $2\times 2$, for which the adjoint Jost functions (rows of the inverse of
 the Jost-function matrix) are directly proportional to Jost functions themselves.
 That is not generic however, and does not hold for the Sasa-Satsuma equation,
 or in general for integrable equations whose spectral operator is $3\times 3$ or higher.

 The derivation of (\ref{deltau}) for the expansion of $(\delta u,
 \delta u^*)^T$ is an important result of this subsection.
 It readily gives the variations in the Sasa-Satsuma fields
 in terms of variations in the initial data. To show this, we first restore
 the time dependence in equation (\ref{deltau}).
 An important fact we need to notice here is that the Jost solutions
 $\Phi$ as defined in (\ref{Phidef}) do not satisfy the time evolution equation of the Lax pair
 of the Sasa-Satsuma equation. Indeed, this time evolution equation
 of the Lax pair is \cite{sasa}
 \be \label{Yt}
 Y_t=-4i\zeta^3\Lambda Y+V(\zeta, u)Y,
 \ee
 where the matrix $V$ goes to zero as $x$ approaches infinity.
 Obviously the large-$x$ asymptotics of $\Phi$ [see (\ref{Jasym}) and
 (\ref{Phidef})] can not satisfy the above equation as $|x|\to \infty$ where $V$ vanishes.
 But this problem can be easily fixed. Defining the ``time-dependent"
 Jost functions
 \be  \label{Phit}
 \Phi^{(t)}=\Phi e^{-4i\zeta^3\Lambda t},
 \ee
 then these functions satisfy {\it both} parts of the Lax pair,
(\ref{spectr}) and (\ref{Yt}). The reason they now satisfy the time
evolution equation (\ref{Yt}) is due to their satisfying the
asymptotic time evolution equation (\ref{Yt}) as $|x|\to \infty$ as
well as the compatibility relation of the Lax pair. Similarly we
define
 the ``time-dependent" adjoint Jost functions as
 \be \label{barPhit}
 \bar{\Phi}^{(t)}=e^{4i\zeta^3\Lambda t}\bar{\Phi},
 \ee
 which satisfy both adjoint equations of the Lax pair. Now we use these new Jost
 functions and adjoint Jost functions to replace those in the definitions (\ref{Z12def}) and
 get ``time-dependent" $(Z_1, Z_2)$ functions
 \be  \label{Z12tdef}
 Z_1^{(t)}=\left[\begin{array}{c}
 \phi_{31}^{(t)}\bar{\phi}_{13}^{(t)}+\phi_{33}^{(t)}\bar{\phi}_{12}^{(t)} \\
 \phi_{32}^{(t)}\bar{\phi}_{13}^{(t)}+\phi_{33}^{(t)}\bar{\phi}_{11}^{(t)}
 \end{array}\right], \quad
 Z_2^{(t)}=\left[\begin{array}{c}
 \phi_{11}^{(t)}\bar{\phi}_{33}^{(t)}+\phi_{13}^{(t)}\bar{\phi}_{32}^{(t)} \\
 \phi_{12}^{(t)}\bar{\phi}_{33}^{(t)}+\phi_{13}^{(t)}\bar{\phi}_{31}^{(t)}
 \end{array}\right].
 \ee
 In view of the relations (\ref{Phit}) and (\ref{barPhit}), we see
 that these ``time-dependent" $(Z_1, Z_2)$ functions are related to
 the original ones as
 \be  \label{Z12t}
 Z_1^{(t)}=Z_1 e^{8i\zeta^3t}, \quad
 Z_2^{(t)}=Z_2 e^{-8i\zeta^3t}.
 \ee

 Another fact we need to notice in conjunction with the
 time-restored equation of (\ref{deltau}) is that
 for the Sasa-Satsuma equation, the time
 evolution of the scattering matrix $S$ is given by \cite{sasa}
 \be  \label{St}
 S_t=-4i\xi^3[\Lambda, S].
 \ee
  Thus
 \be  \label{s33t}
 s'_{33}(t)=0, \quad s'_{31}(t)=8i\xi^3s_{31}(t),
 \ee
 where the prime indicates differentiation with respect to $t$.
 Then recalling the definition (\ref{rou}) of $(\rho_1,
 \bar{\rho}_1)$ as well as the symmetry property (\ref{involS}),
we obtain the time evolution of the varied scattering data
 as
 \be  \label{rho1t}
 \delta \rho_1(\xi, t)=\delta \rho_1(\xi, 0)e^{8i\xi^3 t}, \qquad
 \delta \bar{\rho}_1(\xi, t)=\delta \bar{\rho}_1(\xi, 0)e^{-8i\xi^3 t}.
 \ee

 Now we insert the above time-dependence relations (\ref{Z12t}) and
 (\ref{rho1t}) into the time-restored expansion (\ref{deltau}), and
 get a new expansion
 \be  \label{deltaut}
 \left[\begin{array}{c}
 \delta u(x, t) \\
 \delta u^*(x, t) \end{array}\right]=
 -\frac{1}{\pi} \int_{-\infty}^\infty  \left[Z_1^{(t)}(\xi; x, t) \delta \rho_1(\xi, 0)+Z_2^{(t)}(\xi; x, t)
 \delta \bar{\rho}_1(\xi, 0) \right]d\xi.
 \ee
 The advantages of this expansion are that the Jost functions
 $\Phi^{(t)}$ in $[Z_1^{(t)}, Z_2^{(t)}]$ satisfy both equations of the Lax
 pair, and the expansion coefficients $[\delta\rho_1, \delta\bar{\rho}_1]$
 are time-independent.

 Since $\delta u$ and $\delta u^*$ must satisfy the homogeneous
 linearized Sasa-Satsuma equation, it follows that the functions
 $[Z_1^{(t)}, Z_2^{(t)}]$ appearing in the expansion (\ref{deltaut})
 must also and precisely satisfy the same linearized Sasa-Satsuma equation.
 A similar
 fact for several other integrable equations has been noted before
 by direct verification (see \cite{Benjamin} for instance). Here we
 will give a simple proof of this fact which holds  for any
 integrable equation where an expansion such as (\ref{deltaut}) exists.
 Suppose the linearized operator of the Sasa-Satsuma equation
 for $[u(x, t), u^*(x, t)]^T$ is $\cal{L}$.
 Since the potential $u(x, t)$ satisfies
 the Sasa-Satsuma equation, and the variation of the potential $[\delta u(x, t), \delta u^*(x,
 t)]^T$ is infinitestimal, it follows that
 \be
 {\cal L}\left[\begin{array}{c}
 \delta u(x,t) \\
 \delta u^*(x,t) \end{array}\right]=0.
 \ee
 Inserting Eq. (\ref{deltaut}) into the above equation, we get
 \be \label{Lint}
 \int_{-\infty}^\infty  \left[{\cal L} Z_1^{(t)}(\xi; x, t) \delta \rho_1(\xi, 0)+{\cal L}Z_2^{(t)}(\xi; x, t)
 \delta \bar{\rho}_1(\xi, 0) \right]d\xi=0.
 \ee
 Since the initial variations of the scattering data $(\delta \rho_1,
 \delta\bar{\rho}_1)(\xi, 0)$ are arbitrary and linearly independent,
 it follows that
 \be  \label{LZ1Z2}
 {\cal L} Z_1^{(t)}(\xi; x, t)={\cal L}Z_2^{(t)}(\xi; x, t)=0
 \ee
 for any $\xi\in {\mathbb R}$. Thus
 $Z_1^{(t)}$ and $Z_2^{(t)}$ satisfy the linearized
 Sasa-Satsuma equation and are therefore
 the squared eigenfunctions of this equation.

Now in the derivation of (\ref{deltaut}), from (\ref{RHdF}) forward,
no constraints have been put on the variations of $u$ and $u^*$,
except for the implied assumption that the scattering data for $u$
and $u^*$ and its variations will exist. Whence we expect that
$[\delta u, \delta u^*]^T$ in (\ref{deltaut}) can be considered to
be arbitrary. If so, then due to the equality which we see in
(\ref{deltaut}), we already can expect $\{Z_1^{(t)}(\xi; x,t),
Z_2^{(t)}(\xi; x, t), \xi \in {\mathbb R}\}$ to form a complete set
in an appropriate functional space (such as $L_1$).

We remark that the squared eigenfunctions given here are all
linearly independent. That can be easily verified upon recognizing
that from (\ref{spectr}), one may readily construct an
integro-differential eigenvalue problem, for which these squared
eigenfunctions are the true eigenfunctions. Then in the usual manner
it follows that these squared eigenfunctions are linearly
independent. Similar for the adjoint squared eigenfunctions. The
linear independence of these squared eigenfunctions can also be
easily established after we have obtained the dual relations of
(\ref{deltaut}) which give variations of scattering data due to
arbitrary variations of the potentials (see next section).

 The squared eigenfunctions $[Z_1^{(t)}, Z_2^{(t)}]$ have nice analytic
 properties.
 Indeed, recalling the analytic properties of Jost
 functions discussed in Sec. 2, we see that $Z_1^{(t)}(\zeta) e^{-2i\zeta
 x}$ is analytic in $\mathbb{C}_-$, while $Z_2^{(t)}(\zeta) e^{2i\zeta
 x}$ is analytic in $\mathbb{C}_+$. These analytic properties will
 be essential when we extend our results to the general case
 where $s_{33}$ and $\bar{s}_{33}$ have zeros (see end of this section).

 It should be noted that in the expressions (\ref{Z12tdef}) for $[Z_1^{(t)}, Z_2^{(t)}]$, if
 $\phi_3^{(t)}$ is replaced by other Jost functions (such as $\phi_1^{(t)}$), or
 if $\bar{\phi}_1^{(t)}$ is replaced by other adjoint Jost functions (such
 as $\bar{\phi}_2^{(t)}$), the resulting functions would still satisfy the
 linearized Sasa-Satsuma equation and are thus also squared eigenfunctions of this equation.
 These facts can be verified directly by
 inserting such functions into the linearized Sasa-Satsuma equation and
 noticing that ${\Phi}^{(t)}$ and $\bar{\Phi}^{(t)}$ satisfy the
 Lax pair (\ref{spectr}), (\ref{Yt}) and the
 adjoint Lax pair respectively. Similar results also hold for other integrable equations. However,
 these other squared eigenfunctions in general do not have nice analytic properties in
 ${\mathbb C}_{\pm}$ nor are they linearly independent of the set (\ref{Z12tdef}).

\subsection{Variations of the scattering data and adjoint squared eigenfunctions}

 In this subsection, we calculate variations of the scattering data which occur due to
 arbitrary variations in the potentials. These formulae, together with Eq. (\ref{deltau}) or (\ref{deltaut}),
 will give the ``adjoint" form of the squared eigenfunctions as well as their closure relation.

 We start with the spectral equation (\ref{spectr}) for Jost
 functions $\Phi$, or specifically,
 \be
 \Phi_x=-i\zeta\Lambda \Phi+Q\Phi.   \label{Phi}
 \ee
 Taking the variation to this equation, we get
 \be
 \delta\Phi_x=-i\zeta\Lambda \delta\Phi+Q\delta\Phi+\delta Q \Phi.   \label{dPhi}
 \ee
 Recalling that $\Phi \to e^{-i\zeta \Lambda x}$ as $x\to -\infty$
 [see Eqs. (\ref{Jasym}) and (\ref{Phidef})], thus
 $\delta \Phi \to 0$ as $x\to -\infty$. As a result, the solution of
 the inhomogeneous equation (\ref{dPhi}) can be found by the
 method of variation of parameters as
 \be
 \delta \Phi(\zeta; x)=\Phi(\zeta; x)\int_{-\infty}^x \Phi^{-1}(\zeta;
 y)\delta Q(y) \Phi(\zeta; y)dy.
 \ee
 Here
 \be
 \delta Q=\left(\begin{array}{ccc}0 & 0 & \delta u \\
 0 & 0 & \delta u^* \\
 -\delta u^* & -\delta u & 0 \end{array}\right)
 \ee
 is the variation of the potential.
 Now we take the limit of $x\to \infty$ in the above equation.
 Recalling $\Phi=\Psi S$ and the asymptotics of $\Psi \to
 E$ as $x\to \infty$, we see that $\Phi\to ES$, $\delta\Phi \to E\delta S$
 as $x \to \infty$. Thus in this limit, the above equation becomes
 \be
 \delta S(\xi)=S(\xi) \int_{-\infty}^\infty \Phi^{-1}(\xi;
 x)\delta Q(x) \Phi(\xi; x)dx, \qquad \xi \in \mathbb{R}.
 \ee
 Noticing that $S\Phi^{-1}=\Psi^{-1}\equiv \bar{\Psi}$, the above equation
 can be rewritten as
 \be \label{deltaS}
 \delta S(\xi)=\int_{-\infty}^\infty \bar{\Psi}(\xi;
 x)\delta Q(x) \Phi(\xi; x)dx, \qquad \xi \in \mathbb{R}.
 \ee
 This formula gives variations of scattering coefficients
 $\delta s_{ij}$ via the variation of the potential $\delta Q$.
 From it, the variation of the scattering data $\delta \rho_1$ can be
 readily found. This $\delta \rho_1$ formula contains both Jost
 functions $\Phi$ and adjoint Jost functions $\bar{\Psi}$.
 When we further express $\Phi$ in terms of
 $\Psi$ through the relation (\ref{PhiPsiS})
 (so that the boundary conditions of the Jost functions involved are
 all set at $x=+\infty$), we obtain the expression for $\delta \rho_1$ as
 \be
 \delta\rho_1=\frac{1}{s_{33}^2}\int_{-\infty}^\infty \left[
 \delta u \left( \bar{\psi}_{31} \varphi_3-
 \bar{\psi}_{33}\varphi_2\right)+
 \delta u^* \left(\bar{\psi}_{32}\varphi_3 -
 \bar{\psi}_{33}\varphi_1\right)\right]dx,
 \ee
 where the column vector $\varphi$ is defined as
 \be \label{varphidef}
 \varphi=[\varphi_1, \varphi_2, \varphi_3]^T\equiv
 \bar{s}_{22}\psi_1-\bar{s}_{21}\psi_2.
 \ee
 The variation $\delta \bar{\rho}_1$ can be derived by taking the
 complex conjugate of the above equation and utilizing the symmetry
 relations (\ref{rho_sym}) and (\ref{involphi}). Defining functions
 \be \label{varphidef2}
 \bar{\varphi}=[\bar{\varphi}_1, \bar{\varphi}_2, \bar{\varphi}_3]
 \equiv s_{22}\bar{\psi}_1-s_{12}\bar{\psi}_2,
 \ee
 \be  \label{Omegadef}
 \Omega_1=\left[\begin{array}{c}
 \bar{\psi}_{31}\varphi_3 -\bar{\psi}_{33} \varphi_2 \\
 \bar{\psi}_{32}\varphi_3 -\bar{\psi}_{33} \varphi_1
 \end{array}\right], \quad
 \Omega_2=\left[\begin{array}{c}
 \psi_{32}\bar{\varphi}_3 -\psi_{33} \bar{\varphi}_1\\
 \psi_{31}\bar{\varphi}_3 -\psi_{33} \bar{\varphi}_2
 \end{array}\right],
 \ee
 and inner products
 \be \label{inn_prod}
 \langle f, g \rangle=\int_{-\infty}^\infty f^T(x)g(x)dx,
 \ee
 then the final expressions for variations of the scattering data $\delta \rho_1$
 and $\delta \bar{\rho}_1$ are
 \be \label{drho}
 \delta\rho_1(\xi)=\frac{1}{s_{33}^2(\xi)}\left\langle
 \Omega_1(\xi; x), \; \left[\begin{array}{c}
 \delta u(x) \\
 \delta u^*(x) \end{array}\right]\right\rangle,
 \ee
 \be  \label{drhobar}
 \delta\bar{\rho}_1(\xi)=\frac{1}{\bar{s}_{33}^2(\xi)}\left\langle
 \Omega_2(\xi; x), \; \left[\begin{array}{c}
 \delta u(x) \\
 \delta u^*(x) \end{array}\right]
 \right\rangle,
 \ee
 where $\xi\in \mathbb{R}$.

 The above two equations are associated with the expansion (\ref{deltau})
 of the previous section, where time is frozen in both cases.
 A small problem with these equations is
 that the Jost functions and adjoint Jost functions appearing in the
 definitions (\ref{Omegadef}) of $[\Omega_1, \Omega_2]$ do not
 satisfy the time evolution equation (\ref{Yt}) of the Lax pair (see the previous section).
 To fix this problem, we repeat the practice of the previous section
 and introduce ``time-dependent" versions of
 the functions $[\Omega_1, \Omega_2]$ as
 \be  \label{Omegatdef}
 \Omega_1^{(t)}=\left[\begin{array}{c}
  \bar{\psi}_{31}^{(t)}\varphi_3^{(t)}-\bar{\psi}_{33}^{(t)}\varphi_2^{(t)}  \\
  \bar{\psi}_{32}^{(t)}\varphi_3^{(t)}-\bar{\psi}_{33}^{(t)}\varphi_1^{(t)}
 \end{array}\right], \quad
 \Omega_2^{(t)}=\left[\begin{array}{c}
 \psi_{32}^{(t)}\bar{\varphi}_3^{(t)} -\psi_{33}^{(t)}\bar{\varphi}_1^{(t)} \\
 \psi_{31}^{(t)}\bar{\varphi}_3^{(t)} -\psi_{33}^{(t)}\bar{\varphi}_2^{(t)}
 \end{array}\right],
 \ee
 where $\varphi^{(t)}(\xi, x, t)$ and $\bar{\varphi}^{(t)}(\xi, x, t)$ are defined as
 \be \label{varphitdef}
 \varphi^{(t)}\equiv
 \bar{s}_{22}(\xi)\psi_1^{(t)}-\bar{s}_{21}(\xi)\psi_2^{(t)}, \quad
 \bar{\varphi}^{(t)}
 \equiv s_{22}(\xi)\bar{\psi}_1^{(t)}-s_{12}(\xi)\bar{\psi}_2^{(t)},
 \ee
 $\Phi^{(t)}$ and $\bar{\Phi}^{(t)}$ have been defined in
 (\ref{Phit})-(\ref{barPhit}), and $\Psi^{(t)}$ and
 $\bar{\Psi}^{(t)}$ are defined as
 \be  \label{Psit}
 \Psi^{(t)}=\Psi e^{-4i\zeta^3\Lambda t}, \quad  \bar{\Psi}^{(t)}=e^{4i\zeta^3\Lambda
 t}\bar{\Psi}.
 \ee
 Notice from (\ref{St}) and the symmetry condition (\ref{involS})
 that $(s_{12}, s_{22}, \bar{s}_{21},
 \bar{s}_{22})$ are time-independent.
 Similar to what we have done in the previous section, we can show
 that $\varphi^{(t)}$ and $\Psi^{(t)}$ in the definition
 (\ref{Omegatdef}) are Jost functions satisfying both the Lax pair
 (\ref{spectr}) and (\ref{Yt}), while functions
 $\bar{\varphi}^{(t)}$ and $\bar{\Psi}^{(t)}$ satisfy both the
 adjoint Lax pair. In addition, $[\Omega_1^{(t)}, \Omega_2^{(t)}]$
 are related to $[\Omega_1, \Omega_2]$ as
 \be  \label{Omega12t}
 \Omega_1^{(t)}=\Omega_1 e^{-8i\zeta^3t}, \quad
 \Omega_2^{(t)}=\Omega_2 e^{8i\zeta^3t}.
 \ee
 Inserting this relation and (\ref{rho1t}) into the time-restored
 equations (\ref{drho}) and (\ref{drhobar}), we get the new
 relations
 \be \label{drhot}
 \delta\rho_1(\xi, 0)=\frac{1}{s_{33}^2(\xi)}\left\langle
 \Omega_1^{(t)}(\xi; x, t), \; \left[\begin{array}{c}
 \delta u(x, t) \\
 \delta u^*(x, t) \end{array}\right]\right\rangle,
 \ee
 \be  \label{drhobart}
 \delta\bar{\rho}_1(\xi, 0)=\frac{1}{\bar{s}_{33}^2(\xi)}\left\langle
 \Omega_2^{(t)}(\xi; x, t), \; \left[\begin{array}{c}
 \delta u(x, t) \\
 \delta u^*(x, t) \end{array}\right]
 \right\rangle.
 \ee
 These two new relations are associated with the new expansion
 (\ref{deltaut}) in the previous section.

 The functions $(\Omega_1^{(t)}, \Omega_2^{(t)})$ appearing in the above
 variations-of-scattering-data formulae (\ref{drhot})-(\ref{drhobart}) are precisely adjoint
 squared eigenfunctions satisfying the adjoint linearized Sasa-Satsuma equation.
 Analogous facts for several other integrable equations have been noted
 before by direct verification (see \cite{Benjamin}). Below
 we will present a general proof of this fact which will hold for those integrable equations
 where we have relations similar to (\ref{deltaut}), (\ref{drhot}) and (\ref{drhobart}).
 Along the way, we will also obtain the closure relation, the orthogonality relations and inner
 products between the squared eigenfunctions and the adjoint squared
 eigenfunctions.

 First, we insert Eqs. (\ref{drhot})-(\ref{drhobart}) into
 the expansion (\ref{deltaut}). Exchanging the order of integration,
 we get
 \be
 \left[\begin{array}{c}
 \delta u(x, t) \\
 \delta u^*(x, t) \end{array}\right]=
 -\frac{1}{\pi} \int_{-\infty}^\infty \int_{-\infty}^\infty
 \left[\frac{1}{s_{33}^2(\xi)}Z_1^{(t)}(\xi; x, t)\Omega_1^{(t)T}(\xi; x', t)+
 \frac{1}{\bar{s}_{33}^2(\xi)}Z_2^{(t)}(\xi; x, t)\Omega_2^{(t)T}(\xi;
 x', t)\right]d\xi \left[\begin{array}{c}
 \delta u(x', t) \\
 \delta u^*(x', t) \end{array}\right]dx'.
 \ee
 Since we are considering that $(\delta u, \delta u^*)$ are
arbitrary localized functions (in an appropriate functional space such as
 $L_1$), in order for the above equation to hold, we must have
 \be \label{closure}
 -\frac{1}{\pi} \int_{-\infty}^\infty
 \left[\frac{1}{s_{33}^2(\xi)}Z_1^{(t)}(\xi; x, t)\Omega_1^{(t)T}(\xi; x', t)+
 \frac{1}{\bar{s}_{33}^2(\xi)}Z_2^{(t)}(\xi; x, t)\Omega_2^{(t)T}(\xi;
 x', t)\right]d\xi=\delta(x-x')\openone,
 \ee
 which is the closure relation (discrete-spectrum contributions
 are absent here due to our assumption of $s_{33}$, $\bar{s}_{33}$
 having no zeros). Here $\delta(x)$ is the dirac
 delta function. This closure relation has the usually symmetry in that not only do the
 squared eigenfunctions $\{Z_1^{(t)}(\xi; x, t), Z_2^{(t)}(\xi; x, t), \xi\in \mathbb{R}\}$ form a complete
 set, but also the adjoint functions $\{\Omega_1^{(t)}(\xi; x, t), \Omega_2^{(t)}(\xi; x, t), \xi\in
 \mathbb{R}\}$ form a complete set as well.

To obtain the inner products and orthogonality relations between
these functions, we reverse the above and insert the expansion
(\ref{deltaut}) into Eqs.
 (\ref{drhot})-(\ref{drhobart}). Exchanging the order of integration,
 we get
 \be   \label{rhorho}
 \left[\begin{array}{c}\delta\rho_1(\xi, 0) \\
 \delta\bar{\rho}_1(\xi, 0) \end{array}\right]=\int_{-\infty}^\infty
 M(\xi, \xi', t)\left[\begin{array}{c} \delta\rho_1(\xi', 0) \\
 \delta\bar{\rho}_1(\xi', 0) \end{array}\right]d\xi',
 \ee
 where
 \be
 M(\xi, \xi', t)=\left( \begin{array}{cc} -\frac{1}{\pi s_{33}^2(\xi)}\left\langle \Omega_1^{(t)}(\xi; x, t), Z_1^{(t)}(\xi'; x, t)
 \right\rangle,  & -\frac{1}{\pi s_{33}^2(\xi)}\left\langle \Omega_1^{(t)}(\xi; x, t), Z_2^{(t)}(\xi'; x, t)
 \right\rangle  \\ \\
 -\frac{1}{\pi \bar{s}_{33}^2(\xi)}\left\langle \Omega_2^{(t)}(\xi; x, t), Z_1^{(t)}(\xi'; x, t)
 \right\rangle, & -\frac{1}{\pi \bar{s}_{33}^2(\xi)}\left\langle \Omega_2^{(t)}(\xi; x, t), Z_2^{(t)}(\xi'; x, t)
 \right\rangle \end{array} \right).
 \ee
 Since the relation (\ref{rhorho}) is taken to be valid for arbitrary functions
 of $\delta\rho_1(\xi, 0)$ and $\delta\bar{\rho}_1(\xi, 0)$, thus
 \be
 M(\xi, \xi', t)=\openone \delta(\xi-\xi').
 \ee
 Consequently we get the inner products and orthogonality
 relations between squared eigenfunctions $[Z_1^{(t)}, Z_2^{(t)}]$
 and functions $[\Omega_1^{(t)}, \Omega_2^{(t)}]$ as
 \be  \label{inn1}
 \left\langle \Omega_1^{(t)}(\xi; x, t), Z_1^{(t)}(\xi'; x, t) \right\rangle=-\pi
 s_{33}^2(\xi)\delta(\xi-\xi'),
 \ee
 \be  \label{inn2}
 \left\langle \Omega_2^{(t)}(\xi; x, t), Z_2^{(t)}(\xi'; x, t) \right\rangle=-\pi
 \bar{s}_{33}^2(\xi)\delta(\xi-\xi'),
 \ee
 \be   \label{inn3}
 \left\langle \Omega_1^{(t)}(\xi; x, t), Z_2^{(t)}(\xi'; x, t) \right\rangle=
 \left\langle \Omega_2^{(t)}(\xi; x, t), Z_1^{(t)}(\xi'; x, t) \right\rangle=0
 \ee
 for any $\xi, \xi'\in \mathbb{R}$.

 To show that $(\Omega_1, \Omega_2)$ are adjoint squared eigenfunctions which satisfy the adjoint linearized
 equation of the Sasa-Satsuma equation, we first separate the temporal derivatives from the spatial ones in the
 linearized operator ${\cal L}$ as
 \be \label{Lt}
 {\cal L}=\openone \partial_t +L,
 \ee
 where $L$ only involves spatial derivatives. The adjoint linearized operator is
 then
 \be
 {\cal L}^A=-\openone \partial_t+L^A,
 \ee
 where $L^A$ is the adjoint operator of $L$. Now we take the inner
 product between the equation ${\cal L}Z_1^{(t)}(\xi'; x, t)=0$ and function
 $\Omega_1^{(t)}(\xi; x, t)$. When Eq. (\ref{Lt}) is inserted into it, we find that
 \be  \label{LZ}
 \langle LZ_1^{(t)}(\xi'),\;  \Omega_1^{(t)}(\xi)\rangle- \langle Z_1^{(t)}(\xi'), \;
 \partial_t\Omega_1^{(t)}(\xi)\rangle+\partial_t\langle Z_1^{(t)}(\xi'), \;
 \Omega_1^{(t)}(\xi)\rangle=0.
 \ee
 Here and immediately below, the $(x, t)$ dependence of $Z_{1, 2}^{(t)}$ and $\Omega_{1, 2}^{(t)}$ is suppressed
 for notational simplicity.
 Recall that $s_{33}$ is time-independent for the Sasa-Satsuma
 equation, thus from the inner-product equation (\ref{inn1}) we see
 that $\partial_t\langle Z_1^{(t)}(\xi'), \Omega_1^{(t)}(\xi)\rangle=0$. Using integration
 by parts, the first term in the above equation can be rewritten as
 \be \label{LZ1}
 \langle LZ_1^{(t)}(\xi'), \Omega_1^{(t)}(\xi)\rangle=\langle Z_1^{(t)}(\xi'),
 L^A\Omega_1^{(t)}(\xi)\rangle+W(\xi', \xi)|_{x=-L}^{x=L}, \qquad L \to \infty,
 \ee
 where function $W(\xi', \xi; x, t)$ contains terms which are generated
 during integration by parts. These terms are quadratic combinations
 of $Z_1^{(t)}(\xi')$, $\Omega_1^{(t)}(\xi)$ and their spatial derivatives. When
 $x=\pm L \to \pm \infty$, each term has the form
 $f(\xi')g(\xi)\mbox{exp}(\pm i\xi' L\pm i\xi L\pm 4i\xi'^3t\pm 4i\xi^3t)$ due to the
 large-$x$ asymptotics of Jost functions $\Phi$ and $\Psi$. Here
 $f(\xi')$ and $g(\xi)$ are related to scattering coefficients and are continuous
 functions. Then in the sense of generalized functions, the last
 term in Eq. (\ref{LZ1}) is zero due to the Riemann-Lebesgue lemma.
 Inserting the above results into Eq. (\ref{LZ}), we find that
 \be \label{orth1}
 \langle Z_1^{(t)}(\xi'), {\cal L}^A\Omega_1^{(t)}(\xi)\rangle=0, \qquad \xi', \, \xi\in
 \mathbb{R},
 \ee
 i.e. ${\cal L}^A\Omega_1^{(t)}(\xi)$ is orthogonal to all $Z_1^{(t)}(\xi')$.
 By taking the inner product between the equation ${\cal L}Z_2^{(t)}(\xi')=0$ and function
 $\Omega_1^{(t)}(\xi)$ and doing similar calculations, we readily find that
 \be  \label{orth2}
 \langle Z_2^{(t)}(\xi'), {\cal L}^A\Omega_1^{(t)}(\xi)\rangle=0, \qquad \xi', \; \xi\in
 \mathbb{R}.
 \ee
 In other words, ${\cal L}^A\Omega_1^{(t)}(\xi)$ is also orthogonal to all
 $Z_2^{(t)}(\xi')$.
 Since $\{Z_1^{(t)}(\xi'), Z_2^{(t)}(\xi'), \xi'\in \mathbb{R}\}$ forms a complete
 set, Eqs. (\ref{orth1})-(\ref{orth2}) dictate that ${\cal
 L}^A\Omega_1^{(t)}$ has to be zero for any $\xi, x$ and $t$, i.e.
 \be
 {\cal L}^A\Omega_1^{(t)}(\xi; x, t)=0,  \qquad \xi\in \mathbb{R}.
 \ee
 Thus $\Omega_1^{(t)}(\xi)$ is an adjoint squared eigenfunction
 that satisfies the adjoint linearized Sasa-Satsuma equation for every
 $\xi\in \mathbb{R}$. Similarly, it can be shown that $\Omega_2^{(t)}(\xi)$
 is also an adjoint squared eigenfunction for every
 $\xi\in \mathbb{R}$. In short,
 $\{\Omega_1^{(t)}(\xi), \Omega_2^{(t)}(\xi), \xi\in
 \mathbb{R}\}$ is a complete set of adjoint squared eigenfunctions.

 Like squared eigenfunctions $(Z_1^{(t)}, Z_2^{(t)})$, these adjoint squared
 eigenfunctions $(\Omega_1^{(t)}, \Omega_2^{(t)})$ are also quadratic combinations of Jost
 functions and adjoint Jost functions. In addition, they have nice
 analytic properties as well. To see this latter fact, notice from the
 definition (\ref{varphidef}) that $\varphi e^{i\zeta x}$ is
 analytic in $\mathbb{C}_-$ since $\bar{s}_{21}, \bar{s}_{22}, \psi_1 e^{i\zeta
 x}$ and $\psi_2 e^{i\zeta x}$ are analytic in $\mathbb{C}_-$.
 Then recalling that $\bar{\psi}_3e^{i\zeta x}$ is also analytic in $\mathbb{C}_-$,
 we see that $\Omega_1^{(t)}(\zeta) e^{2i\zeta x}$ is
 analytic in $\mathbb{C}_-$. Similarly, $\Omega_2^{(t)}(\zeta) e^{-2i\zeta x}$ is
 analytic in $\mathbb{C}_+$.

 It is worthy to point out that in the above adjoint squared eigenfunctions
 (\ref{Omegatdef}), if $\varphi^{(t)}$ is replaced by the Jost function
 $\psi_1^{(t)}$ (or $\psi_2^{(t)}$), or if $\bar{\varphi}^{(t)}$ is replaced by the adjoint Jost function
 $\bar{\psi}_1^{(t)}$ (or $\bar{\psi}_2^{(t)}$), the resulting functions would still
 satisfy the adjoint linearized Sasa-Satsuma equation. Thus one may be tempted to
 simply take $\varphi^{(t)}=\psi_k^{(t)}$, $\bar{\varphi}^{(t)}=\bar{\psi}_k^{(t)}$ ($k=1$ or
 2) rather than (\ref{varphitdef}) as adjoint squared
 eigenfunctions. The problem with these ``simpler" adjoint squared eigenfunctions is that
 their inner products with squared eigenfunctions (\ref{Z12tdef})
 are not what they are supposed to be [see
 (\ref{inn1})-(\ref{inn3})], neither does one get the closure
 relation (\ref{closure}). Thus the adjoint squared eigenfunctions
 (\ref{Omegatdef}) coming from our systematic procedure of variation calculations are the
 correct ones to use.

\subsection{Extension to the general case}
 In the previous subsections, the squared eigenfunctions and their
 closure relation were established under the assumption that $s_{33}$
 and $\bar{s}_{33}$ have no zeros, i.e. the spectral equation
 (\ref{spectr}) has no discrete eigenvalues. In this subsection, we
 extend those results to the general case where $s_{33}$ and
 $\bar{s}_{33}$ have zeros in their respective planes of analyticity
 $\mathbb{C}_{\pm}$. Due to the symmetry properties (\ref{involS})
 and (\ref{Saddition}), we have
 \be \label{s33sym}
 \bar{s}_{33}(\zeta)=s_{33}^*(\zeta^*), \quad
 s_{33}(\zeta)=s_{33}^*(-\zeta^*).
 \ee
 Thus if $\zeta_j \in
 \mathbb{C}_{-}$ is a zero of $s_{33}$, i.e. $s_{33}(\zeta_j)=0$,
 then $-\zeta_j^* \in \mathbb{C}_{-}$ is also a zero of $s_{33}$, and
 $\{\zeta_j^*, -\zeta_j\}\in \mathbb{C}_{+}$ are both zeros of
 $\bar{s}_{33}$. This means that zeros of $s_{33}$ and $\bar{s}_{33}$
 always appear in quadruples. Suppose all the zeros of $s_{33}$ and
 $\bar{s}_{33}$ are $\zeta_j \in \mathbb{C}_{-}$ and $\bar{\zeta}_j
 \in \mathbb{C}_{+}$, $j=1, ..., 2N$, respectively. For simplicity,
 we also assume that all zeros are simple.

 In this general case with zeros, squared eigenfunctions $Z_1^{(t)}, Z_2^{(t)}$
 clearly still satisfy the linearized Sasa-Satsuma equation, and
 adjoint squared eigenfunctions $\Omega_1^{(t)}, \Omega_2^{(t)}$ still satisfy
 the adjoint linearized Sasa-Satsuma equation --- facts which
 will not change when $s_{33}$ and $\bar{s}_{33}$ have zeros. The
 main difference from the previous no-zero case is that, the sets of
 (continuous) squared eigenfunctions $\{Z_1^{(t)}(\xi; x), Z_2^{(t)}(\xi; x),
 \xi\in \mathbb{R}\}$ and adjoint squared eigenfunctions
 $\{\Omega_1^{(t)}(\xi; x), \Omega_2^{(t)}(\xi; x), \xi\in \mathbb{R}\}$ are no
 longer complete, i.e. the closure relation (\ref{closure}) does not
 hold any more, and contributions from the discrete spectrum must be
 included now. To account for discrete-spectrum contributions,
 one could proceed by adding variations in the discrete
 scattering data in the above derivations. But a much easier way is to
 simply pick up the pole contributions to the integrals in the closure
 relation (\ref{closure}), as we will do below.
 Recall from the previous subsections that
 $Z_1^{(t)}(\zeta; x, t) e^{-2i\zeta x}$ and $\Omega_1^{(t)}(\zeta; x, t)e^{2i\zeta x}$
 are analytic in $\mathbb{C}_-$, while $Z_2^{(t)}(\zeta; x, t) e^{2i\zeta x}$
 and $\Omega_2^{(t)}(\zeta; x, t)e^{-2i\zeta x}$ are analytic in
 $\mathbb{C}_+$. In addition, from the large-$\zeta$ asymptotics
 (\ref{sijasym1})-(\ref{sijasym2}), (\ref{normal}), and symmetry relations
 (\ref{involphi}), we easily find that
 \be
 Z_1^{(t)}(\zeta; x, t) \to e^{2i\zeta x+8i\zeta^3t} (0, 1)^T, \qquad Z_2^{(t)}(\zeta; x, t)\to
 e^{-2i\zeta x-8i\zeta^3t} (1, 0)^T, \qquad \zeta \to \infty,
 \ee
 \be
 \Omega_1^{(t)}(\zeta; x, t) \to -e^{-2i\zeta x-8i\zeta^3t} (0, 1)^T, \qquad \Omega_2^{(t)}(\zeta;
 x, t) \to -e^{2i\zeta x+8i\zeta^3t} (1, 0)^T, \qquad \zeta \to \infty.
 \ee
 Thus,
 \be
 \int_{C^-}
 \frac{1}{s_{33}^2(\zeta)}Z_1^{(t)}(\zeta; x, t)\Omega_1^{(t)T}(\zeta; x', t)d\zeta
 =-\int_{C^-} e^{2i\zeta (x-x')}d\zeta\; \mbox{diag}(0, 1),
 \ee
 where the integration path $C^-$ is the lower semi-circle of infinite
 radius in anti-closewise direction. Since the function $e^{2i\zeta (x-x')}$ in the
 right hand side of the above equation is analytic, its integration path
 can be brought up to the real axis $\zeta\in \mathbb{R}$. In the
 sense of generalized functions, that integral is equal to $\pi
 \delta(x-x')$, thus
 \be
 \int_{C^-}
 \frac{1}{s_{33}^2(\zeta)}Z_1^{(t)}(\zeta; x, t)\Omega_1^{(t)T}(\zeta; x', t)d\zeta
 =-\pi\delta(x-x') \mbox{diag}(0, 1).
 \ee
 Doing the same calculation for the integral of
 $Z_2^{(t)}\Omega_2^{(t)T}/\bar{s}_{33}^2$ along the upper semi-circle of infinite
 radius $C^+$ in closewise direction and combining it with the above equation,
 we get
 \be
 -\frac{1}{\pi}\int_{C^-}
 \frac{1}{s_{33}^2(\zeta)}Z_1^{(t)}(\zeta; x, t)\Omega_1^{(t)T}(\zeta; x', t)d\zeta
 -\frac{1}{\pi}\int_{C^+}
 \frac{1}{\bar{s}_{33}^2(\zeta)}Z_2^{(t)}(\zeta; x, t)\Omega_2^{(t)T}(\zeta; x', t)d\zeta
 =\delta(x-x') \openone.
 \ee
 Now we evaluate the two integrals in the above equation by bringing
 down the integration paths to the real axis and picking up pole
 contributions by the residue theorem. Recalling that zeros of
 $s_{33}$ and $\bar{s}_{33}$ are simple, the poles in the above
 integrals are second-order. After simple calculations, we get
 the closure relation for the general case (with discrete-spectrum
 contributions) as:
 \begin{eqnarray} \label{closure2}
 -\frac{1}{\pi} \int_{-\infty}^\infty
 \left[\frac{1}{s_{33}^2(\xi)}Z_1^{(t)}(\xi; x, t)\Omega_1^{(t)T}(\xi; x', t)+
 \frac{1}{\bar{s}_{33}^2(\xi)}Z_2^{(t)}(\xi; x, t)\Omega_2^{(t)T}(\xi;
 x', t)\right]d\xi   & \nonumber  \\
 -\sum_{j=1}^{2N} \frac{2i}{s_{33}^{'2}(\bar{\zeta}_j)}\left[Z_1^{(t)}(\bar{\zeta}_j;
 x, t)\Theta_{1}^{(t)T}(\bar{\zeta}_j; x', t)+\frac{\partial Z_1^{(t)}}{\partial \zeta}(\bar{\zeta}_j;
 x, t)\Omega_1^{(t)T}(\bar{\zeta}_j; x', t)\right]  &  \nonumber  \\
 +\sum_{j=1}^{2N} \frac{2i}{\bar{s}_{33}^{'2}({\zeta}_j)}\left[Z_2^{(t)}({\zeta}_j;
 x, t)\Theta_{2}^{(t)T}({\zeta}_j; x', t)+\frac{\partial Z_2^{(t)}}{\partial \zeta}({\zeta}_j;
 x, t)\Omega_2^{(t)T}({\zeta}_j; x', t)\right] &
 =\delta(x-x')\openone,
 \end{eqnarray}
 where
 \be
 \Theta_{1}^{(t)}(\bar{\zeta}_j; x, t)=\frac{\partial \Omega_1^{(t)}}{\partial \zeta}(\bar{\zeta}_j;
 x, t)-\frac{s_{33}^{''}(\bar{\zeta}_j)}{s_{33}^{'}(\bar{\zeta}_j)}\Omega_1^{(t)}(\bar{\zeta}_j;
 x, t),
 \ee
 \be
 \Theta_{2}^{(t)}({\zeta}_j; x, t)=\frac{\partial \Omega_2^{(t)}}{\partial \zeta}({\zeta}_j;
 x, t)-\frac{\bar{s}_{33}^{''}({\zeta}_j)}{\bar{s}_{33}^{'}({\zeta}_j)}
 \Omega_2^{(t)}({\zeta}_j; x, t).
 \ee
which have been termed ``derivative states" \cite{Kaup1976a, Kaup1976b} due to the
differentiation with respect to $\zeta$.
 Clearly, $Z_1^{(t)}(\bar{\zeta}_j)$, $\partial Z_1^{(t)}/\partial \zeta \:(\bar{\zeta}_j)$,
 $Z_2^{(t)}({\zeta}_j)$, and $\partial Z_2^{(t)}/\partial \zeta \: ({\zeta}_j)$ satisfy the linearized Sasa-Satsuma
 equation and are thus discrete squared
 eigenfunctions, while $\Omega_1^{(t)}(\bar{\zeta}_j)$, $\Theta_{1}^{(t)}(\bar{\zeta}_j)$,
 $\Omega_2^{(t)}({\zeta}_j)$, and $\Theta_{2}^{(t)}({\zeta}_j)$ satisfy the adjoint linearized Sasa-Satsuma equation
 and are thus discrete adjoint squared eigenfunctions.

 In soliton perturbation theories, explicit expressions for squared eigenfunctions
 under soliton potentials are needed. For such potentials,
 $s_{31}=s_{32}=\bar{s}_{13}=\bar{s}_{23}=0$, thus $G=\openone$ in the Riemann-Hilbert problem
 (\ref{RHP}). Solutions of this Riemann-Hilbert problem have been
 solved completely for any number of zeros and any orders of their
 algebraic and geometric multiplicities \cite{Shch_Yang}. In the
 simplest case where all zeros of the solitons are simple, solutions
 $P_\pm$ can be found in Eq. (76) of \cite{Shch_Yang} (setting
 $r^{(i}=r^{(j)}=1$). One only needs to keep in mind here that the zeros of
 $s_{33}$ and $\bar{s}_{33}$ always appear in quadruples due to the
 symmetries (\ref{s33sym}). Once solutions $P_\pm$ are obtained, the
 solitons $u(x, t)$ follow from Eq. (\ref{reconst}). Taking the
 large-$x$ asymptotics of $P_\pm$ and utilizing the symmetry
 conditions (\ref{involS}), the full scattering matrix $S$ can be
 readily obtained. Then together with $P_\pm$, explicit expressions for
 all the Jost functions
 $\Phi$, $\Psi$ as well as squared eigenfunctions are then
 derived.

\section{Summary and discussion}
In this paper, squared eigenfunctions were derived for the
Sasa-Satsuma equation. It was shown that these squared
eigenfunctions are sums of two terms, where each term is a product
of a Jost function and an adjoint Jost function. The procedure of
this derivation consists of two steps: one is to calculate the
variations of the potentials via variations of the scattering data
by the Riemann-Hilbert method. The other one is to calculate
variations of the scattering data via variations of the potentials
through elementary calculations. It was proved that the functions
appearing in these variation relations are precisely the squared
eigenfunctions and adjoint squared eigenfunctions satisfying
respectively the linearized equation and the adjoint linearized
equation of the Sasa-Satsuma system. More importantly, our proof is
quite general and generally holds for other integrable equations.
Since the spectral operator of the Sasa-Satsuma equation is $3\times
3$, while its linearized operator is $2\times 2$, we demonstrated
how the two-component squared eigenfunctions are built from only
selected components of the three-component Jost functions, and we
have seen that symmetry properties of Jost functions and the
scattering data play an important role here.

The derivation used in this paper for squared eigenfunctions is a
universal technique. Our main contributions to this method are
triple-fold. First, we showed that for a general integrable system,
if one can obtain the variational relations between the potentials
and the scattering data in terms of the components of the Jost
functions and their adjoints, then from these variational relations
one can immediately identify the squared eigenfunctions and adjoint
squared eigenfunctions which satisfy respectively the linearized
integrable equation and its adjoint equation. Second, we showed that
from these variational relations, one can read off the values of
nonzero inner products between squared eigenfunctions and adjoint
squared eigenfunctions. Thus no longer is it necessary to calculate
these inner products from the Wronskian relations and the
asymptotics of the Jost functions. Thirdly, we showed that from
these variational relations, one can immediately obtain the closure
relation of squared eigenfunctions and adjoint squared
eigenfunctions. After squared eigenfunctions of an integrable
equation are obtained, one then can readily derive the recursion
operator for that integrable equation, since the squared
eigenfunctions are eigenfunctions of the recursion operator
\cite{AKNS,Fokas_commu_kdv, YangJMP,YangStud}.

An important remark about this derivation of squared eigenfunctions
is that it uses primarily the spectral operator of the Lax pair.
Indeed, the key variation relations (\ref{deltau}), (\ref{drho}) and
(\ref{drhobar}) were obtained exclusively from the spectral operator
(\ref{spectr}), while the ``time-dependent" versions of these
relations (\ref{deltaut}), (\ref{drhot}) and (\ref{drhobart}) were
obtained by using the asymptotic form of the time-evolution equation
in the Lax pair as $|x|\to \infty$. This means that all integrable
equations with the same spectral operator would share the same
squared eigenfunctions (except an exponential-in-time factor which
is equation-dependent). For instance, a whole hierarchy of
integrable equations would possess the ``same" squared
eigenfunctions, since the spectral operators of a hierarchy are the
same. This readily reproduces earlier results in
\cite{Fokas_commu_kdv, YangJMP,YangStud}, where such results were
obtained by the commutability relations between the linearized
operators and the recursion operator of a hierarchy.

Since the derivation in this paper uses primarily the spectral
operator of the Lax pair, if two integrable equations have similar
spectral operators, derivations of their squared eigenfunctions
would be similar. For instance, the spectral operator of the Manakov
equations is very similar to that of the Sasa-Satsuma equation
(except that the potential term $Q$ in the Sasa-Satsuma spectral
operator (\ref{spectr}) possesses one more symmetry). Making very
minor modifications to the calculations for the Sasa-Satsuma
equation in the text, we can obtain squared eigenfunctions for the
Manakov equations. This will be demonstrated in the Appendix.

With the squared eigenfunctions obtained for the Sasa-Satsuma
equation, a perturbation theory for Sasa-Satsuma solitons can now be
developed. This problem lies outside the scope of the present
article, and will be left for future studies.

\section*{Acknowledgment}
We thank Dr. Taras Lakoba for very helpful discussions. The work of
J.Y. was supported in part by the Air Force Office of Scientific
Research under grant USAF 9550-05-1-0379.  The work of D.J.K. has
been support in part by the US National Science Foundation under
grant DMS-0505566.

\section*{Appendix: Squared eigenfunctions for the Manakov equation}
 The Manakov equations are \cite{manakov}
 \be
 iu_t+u_{xx}+2(|u|^2+|v|^2)u=0,
 \ee
 \be
 iv_t+v_{xx}+2(|u|^2+|v|^2)v=0.
 \ee
 The spectral operator for the Manakov equations is Eq. (\ref{spectr})
 with
 \be  \label{Q2}
 Q=\left(\begin{array}{ccc}0 & 0 & u \\
 0 & 0 & v \\
 -u^* & -v^* & 0 \end{array}\right).
 \ee
 Below we will use the same notations on Jost functions and the
 scattering matrix as in the main text. Since $Q$ above is also
 anti-Hermitian, Jost functions and the scattering matrix satisfy
 the involution properties (\ref{involJ}) and (\ref{involS}) as
 well. The main difference between the Sasa-Satsuma equation and the
 Manakov equations is that, the potential $Q$ above for the Manakov equations
 does not possess the additional symmetry (\ref{Qsym}), thus the
 Jost functions and the scattering matrix do not possess the
 symmetries (\ref{Jaddition}) and (\ref{Saddition}). In addition,
 the asymptotics (\ref{Fasym}) also needs minor modification.
 It is noted that the
 spectral operator of the Manakov equations is $3\times 3$, while
 the linearized operator of these equations is $4\times 4$. So
 one needs to build four-component squared eigenfunctions from
 three-component Jost functions. This contrasts the Sasa-Satsuma
 equation where one builds two-component squared eigenfunctions from
 three-component Jost functions.

 Repeating the same calculations as in the main text with the above
 minor modifications kept in mind, we still get Eq. (\ref{du}) for
 the variations $(\delta u,
 \delta u^*)$, while the expressions for variations $(\delta v, \delta
 v^*)$ are Eq. (\ref{du}) with $\Pi_{13}$ and $\Pi_{31}$ replaced by
 $\Pi_{23}$ and $\Pi_{32}$. The expression for matrix $\Pi$ is still
 given by Eq. (\ref{Pi}). Thus variations of the potentials via variations
 of the scattering data are found to be
 \be
 [\delta u, \delta v, \delta u^*, \delta v^*]^T=
 -\frac{1}{\pi} \int_{-\infty}^\infty  \left[Z_1(\xi; x) \delta \rho_1(\xi)+Z_2(\xi; x)
 \delta \rho_2(\xi) + Z_3(\xi; x) \delta \bar{\rho}_1(\xi) +Z_4(\xi; x)
 \delta \bar{\rho}_2(\xi)\right]d\xi,
 \ee
 where
 \be  \label{Z1234}
 Z_1=\left[\begin{array}{c}
 \phi_{31}\bar{\phi}_{13}\\
 \phi_{32}\bar{\phi}_{13}\\
 \phi_{33}\bar{\phi}_{11}\\
 \phi_{33}\bar{\phi}_{12}
 \end{array}\right], \quad
 Z_2=\left[\begin{array}{c}
 \phi_{31}\bar{\phi}_{23}\\
 \phi_{32}\bar{\phi}_{23}\\
 \phi_{33}\bar{\phi}_{21}\\
 \phi_{33}\bar{\phi}_{22}
 \end{array}\right], \quad
 Z_3=\left[\begin{array}{c}
 \phi_{11}\bar{\phi}_{33}\\
 \phi_{12}\bar{\phi}_{33}\\
 \phi_{13}\bar{\phi}_{31}\\
 \phi_{13}\bar{\phi}_{32}
 \end{array}\right], \quad
 Z_4=\left[\begin{array}{c}
 \phi_{21}\bar{\phi}_{33}\\
 \phi_{22}\bar{\phi}_{33}\\
 \phi_{23}\bar{\phi}_{31}\\
 \phi_{23}\bar{\phi}_{32}
 \end{array}\right].
 \ee
 Variations of the scattering matrix $\delta S$ are still given by
 Eq. (\ref{deltaS}), with $\delta Q$ modified in view of the form of
 (\ref{Q2}). Then defining vectors
 \be
 \varphi_1=(\varphi_{1j})\equiv \bar{s}_{22}\psi_1-\bar{s}_{21}\psi_2, \quad
 \varphi_2=(\varphi_{2j})\equiv -\bar{s}_{12}\psi_1+\bar{s}_{11}\psi_2,
 \ee
 \be
 \bar{\varphi}_1=(\bar{\varphi}_{1j})
 \equiv s_{22}\bar{\psi}_1-s_{12}\bar{\psi}_2, \quad
 \bar{\varphi}_2=(\bar{\varphi}_{2j})
 \equiv -s_{21}\bar{\psi}_1+s_{11}\bar{\psi}_2,
 \ee
 we find that variations of the scattering data are
 \be
 \delta\rho_1(\xi)=\frac{1}{s_{33}^2(\xi)}\left\langle
 \Omega_1(\xi; x), \; \left[\begin{array}{c}
 \delta u(x) \\
 \delta v(x) \\
 \delta u^*(x) \\
 \delta v^*(x) \end{array}\right]\right\rangle, \quad
 \delta\rho_2(\xi)=\frac{1}{s_{33}^2(\xi)}\left\langle
 \Omega_2(\xi; x), \; \left[\begin{array}{c}
 \delta u(x) \\
 \delta v(x) \\
 \delta u^*(x) \\
 \delta v^*(x) \end{array}\right]\right\rangle,
 \ee

 \be
 \delta\bar{\rho}_1(\xi)=\frac{1}{\bar{s}_{33}^2(\xi)}\left\langle
 \Omega_3(\xi; x), \; \left[\begin{array}{c}
 \delta u(x) \\
 \delta v(x) \\
 \delta u^*(x) \\
 \delta v^*(x)
 \end{array}\right]
 \right\rangle, \quad
 \delta\bar{\rho}_2(\xi)=\frac{1}{\bar{s}_{33}^2(\xi)}\left\langle
 \Omega_4(\xi; x), \; \left[\begin{array}{c}
  \delta u(x) \\
 \delta v(x) \\
 \delta u^*(x) \\
 \delta v^*(x)
 \end{array}\right]
 \right\rangle,
 \ee
 where
 \be  \label{Omega1234}
 \Omega_1=\left[\begin{array}{r}
 \bar{\psi}_{31}\varphi_{13}\\
 \bar{\psi}_{32}\varphi_{13} \\
 -\bar{\psi}_{33}\varphi_{11} \\
 -\bar{\psi}_{33}\varphi_{12}
 \end{array}\right], \quad
 \Omega_2=\left[\begin{array}{r}
 \bar{\psi}_{31}\varphi_{23}\\
 \bar{\psi}_{32}\varphi_{23} \\
 -\bar{\psi}_{33}\varphi_{21} \\
 -\bar{\psi}_{33}\varphi_{22}
 \end{array}\right], \quad
 \Omega_3=\left[\begin{array}{r}
 -\bar{\varphi}_{11}{\psi}_{33} \\
 -\bar{\varphi}_{12}{\psi}_{33} \\
 \bar{\varphi}_{13}{\psi}_{31} \\
 \bar{\varphi}_{13}{\psi}_{32}
 \end{array}\right], \quad
 \Omega_4=\left[\begin{array}{r}
 -\bar{\varphi}_{21}{\psi}_{33} \\
 -\bar{\varphi}_{22}{\psi}_{33} \\
 \bar{\varphi}_{23}{\psi}_{31}\\
 \bar{\varphi}_{23}{\psi}_{32}
 \end{array}\right].
 \ee
 For Manakov equations, the time evolution operator of the Lax pair
 is
 \be
 Y_t=-2i\zeta^2\Lambda Y+V(\zeta, u)Y,
 \ee
 where $V$ goes to zero as $x$ approaches infinity.
 Introducing ``time-dependent" Jost functions and adjoint Jost
 functions such as
 \be
 \Phi^{(t)}=\Phi e^{-2i\zeta^2\Lambda t},
 \quad \bar{\Phi}^{(t)}=e^{2i\zeta^2\Lambda t}\bar{\Phi},
 \ee
 and use them to replace the original Jost functions and adjoint
 Jost functions in the $Z_k$ and $\Omega_k$ definitions (\ref{Z1234}) and
 (\ref{Omega1234}), the resulting functions $Z^{(t)}_{1, 2, 3, 4}$
 and $\Omega^{(t)}_{1, 2, 3, 4}$ would be squared eigenfunctions and
 adjoint squared eigenfunctions of the Manakov equations.
 Equivalently,
 $\{Z_1e^{-4i\zeta^2t}, Z_2e^{-4i\zeta^2t}, Z_3e^{4i\zeta^2t}, Z_4e^{4i\zeta^2t}\}$ are squared eigenfunctions, and
 $\{\Omega_1e^{4i\zeta^2t}, \Omega_2e^{4i\zeta^2t}, \Omega_3e^{-4i\zeta^2t}, \Omega_4e^{-4i\zeta^2t}\}$ adjoint squared
 eigenfunctions, of the Manakov equations.

 It is noted that squared eigenfunctions for the Manakov equations have
 been given in \cite{Gerdjikov} under different notations and derivations.
 Our expressions above are more explicit and easier to use.

\end{document}